%% file: paper.tex
\documentclass[sigconf,authorversion,nonacm]{acmart}

\usepackage{cleveref}
\usepackage{graphicx}
\usepackage[utf8]{inputenc}
\usepackage[caption=false]{subfig}
\usepackage{color}
\usepackage{url}
\usepackage[T1]{fontenc}
\usepackage{cleveref}
\usepackage{blindtext}
\usepackage{booktabs}
\usepackage{enumitem}
\usepackage[title]{appendix}
\usepackage{stfloats}
\usepackage{titlesec}
\usepackage{balance}
\usepackage[font=small,format=plain,labelfont=bf,textfont=it]{caption}
\crefformat{section}{\S#2#1#3}

\AtBeginDocument{%
  \providecommand\BibTeX{{%
    \normalfont B\kern-0.5em{\scshape i\kern-0.25em b}\kern-0.8em\TeX}}}

\usepackage[all=normal,wordspacing]{savetrees}

\title{Characterizing Service Provider Response to the COVID-19 Pandemic in
the United States}

\author{Shinan Liu}
\email{shinanliu@uchicago.edu}
\affiliation{%
  \institution{The University of Chicago}
}

\author{Paul Schmitt}
\email{pschmitt@cs.princeton.edu}
\affiliation{%
  \institution{Princeton University}
}

\author{Francesco Bronzino}
\email{francesco.bronzino@univ-smb.fr}
\affiliation{%
  \institution{Université Savoie Mont Blanc}
}

\author{Nick Feamster}
\email{feamster@uchicago.edu}
\affiliation{%
  \institution{The University of Chicago}
}

\renewcommand{\shortauthors}{Liu, et al.}

\input{abstract}


\begin{document}

\settopmatter{printfolios=true}
\maketitle

\input{introduction}
\input{background}
\input{data}

\input{traffic}
\input{performance}

\input{capacity}
\label{lastpage}\input{conclusion}

\pagebreak
\balance
\bibliographystyle{ACM-Reference-Format}
\bibliography{paper}

\newpage
\appendix
\input{appendix}

\end{document}

%% file: abstract.tex
\begin{abstract}
    The COVID-19 pandemic has resulted in dramatic changes to the daily habits
    of billions of people. Users increasingly have to rely on home broadband
    Internet access for work, education, and other activities. These changes
    have resulted in corresponding changes to Internet traffic patterns. This
    paper aims to characterize the effects of these changes with respect to
    Internet service providers in the United States. We study three questions:
    (1)~How did traffic demands change in the United States as a result of the
    COVID-19 pandemic?; (2)~What effects have these changes had on Internet
    performance?; (3)~How did service providers respond to these changes? We
    study these questions using data from a diverse collection of sources. Our
    analysis of interconnection data for two large ISPs in the United States
    shows a 30--60\% increase in peak traffic rates in the first quarter of
    2020. In particular, we observe traffic downstream peak volumes for a major
    ISP increase of 13--20\% while upstream peaks increased by more than 30\%.
    Further, we observe significant variation in performance across ISPs in
    conjunction with the traffic volume shifts, with evident latency increases
    after stay-at-home orders were issued, followed by a stabilization of
    traffic after April. Finally, we observe that in response to changes in
    usage, ISPs have aggressively augmented capacity at interconnects, at more
    than twice the rate of normal capacity augmentation. Similarly, video
    conferencing applications have increased their network footprint, more than
    doubling their advertised IP address space.
\end{abstract}

%% file: introduction.tex
\section{Introduction}\label{sec:introduction}

The COVID-19 pandemic has resulted in dramatic shifts in the behavioral patterns
of billions of people. These shifts have resulted in corresponding changes in
how people use the Internet. Notably, people are increasingly reliant on home
broadband Internet access for work, education, and other activities. The changes
in usage patterns have resulted in corresponding changes in network traffic
demands observed by Internet service providers. Many reports have noted some of
the effects of these changes from service provider networks~\cite{comcast2020covid,
att2020our}, application providers~\cite{nokia2020network,sandvine2020covid}, 
and Internet exchange points~\cite{oecd2020keeping}. Generally, previous
findings and conventional wisdom suggest that while overall traffic demands
increased, the Internet responded well in response to these changing demands. 

Previous work has shed light on the nature of the resulting changes in traffic
patterns. In Europe, Internet exchange points saw a 15--20\% increase in overall
traffic volumes~\cite{candela2020impact}, in some cases resulting in peaks in
round trip latency in some countries (e.g., Italy) that were approximately 30\%
higher than normal~\cite{feldmann2020lockdown}. Because users were less mobile,
downlink traffic volume decreased by up to 25\% for cellular networks in the
UK~\cite{lutu2020characterization}. While some of the characteristics of
shifting traffic demands are known, and certain aspects of the Internet's
resilience in the face of the traffic shifts are undoubtedly a result of robust
design of the network and protocols, some aspects of the Internet's resilience
are a direct result of providers' swift responses to these changing traffic
patterns. This paper explores these traffic effects from a longitudinal
perspective---exploring traffic characteristics during the first half of 2020 to
previous years---and also explores how service providers {\em responded} to the
changes in traffic patterns.

Service providers and regulatory agencies implemented various responses to the
traffic shifts resulting from COVID-19. AT\&T and Comcast have made public
announcements about capacity increases in response to increases in network
load~\cite{att2020our,comcast2020covid}. The Federal Communications Commission
(FCC) also announced the ``Keep Americans Connected'' initiative to grant
providers (such as AT\&T, Sprint, T-Mobile, U.S. Cellular, Verizon, and others)
additional spectrum to support increased broadband usage~\cite{fcc2020keep}. Web
conferencing applications Zoom and WebEx were also granted temporary relief from
regulatory actions~\cite{fcc2020keep}. These public documents provide some
perspectives on responses, but to date, there are few independent reports and
studies of provider responses. This paper provides an initial view into how some
providers responded in the United States.

We study the effects of the shifts in Internet traffic resulting from the
COVID-19 pandemic response on Internet infrastructure. We study three questions:
\begin{itemize}
    \item \textit{How did traffic patterns change as a result of COVID-19?}
    Traffic volumes and network utilization are changing as a reaction to
    changes in user behaviors. It is critical to measure the exact
    alterations in a long time span.
    \item \textit{What were the resulting effects on performance?} Considering
    an expected surge around the dates when states issued stay-at-home orders or
    declared states of emergency, we seek to observe possible changes in the
    latency and throughput of network traffic across locations. Further,
    different ISPs also have different capacity and provisioning strategies,
    which provides us a finer granularity based on these differences.
    \item \textit{How did ISPs and service providers respond?} Finally, to deal
    with the usage boosts and performance degradations during the COVID-19
    response, operations and reactions of ISPs and service providers were taken
    which may explain the changes in network performance. The answer to this
    question informs us of the networks robustness and their effective disaster
    provisioning strategies. These questions have become increasingly critical
    during the COVID-19 pandemic, as large fractions of the population have come
    to depend on reliable Internet access that performs well for a variety of
    applications, from video conferencing to remote learning and healthcare. 
\end{itemize}
\noindent
To answer these questions, we study a diverse collection of datasets about
network traffic load, through granular measurements, proprietary
data sharing agreements, and user experiences, as well as extensive baseline
data spanning over two years. 

\paragraph{Summary of findings.} First, we study the traffic pattern changes in
the United States (\cref{sec:traffic}) and find that, similar to the changes
previously explored for European networks, our analysis reveals a 30--60\%
increase in peak traffic volumes. In the Comcast network in particular, we find
that downstream peak traffic volume increased 13--20\%, while upstream peak
traffic volume increases by more than 30\%. Certain interconnect peers exhibit
significant changes in the magnitude of traffic during the lockdown. Second, we
observe a temporary, statistically significant increase in latency lasting
approximately two months (\cref{sec:performance}). We observe a temporary
increase of about 10\% in average latency around the time that stay-at-home
orders were issued. Typical latency values returned to normal a few months after
these orders were put in place. We also find heterogeneity between different
ISPs. Finally, we explore how service providers responded to this increase in
traffic demands by adding capacity (\cref{sec:capacity}). ISPs aggressively
added capacity at interconnects, more than 2x the usual rates. On a similar
note, application service providers (e.g., video conferencing apps) increased
the advertised IP address space by 2.5--5x to cope with the corresponding 2--3x
increase in traffic demand.

%% file: background.tex
\section{Related Work}\label{sec:related} 

The pandemic response has modified people's habits, causing them to rely heavily
on the Internet for remote work, e-learning, video streaming, etc. In this
section, we present some previous efforts in measuring the effects of COVID-19 and
past disaster responses on networks and applications.

\paragraph{Network Measurements during COVID-19.}
Previous work has largely focused on aggregate traffic statistics surrounding
the initial COVID-19 lockdowns. Traffic surged about 20\% in Europe for
broadband networks~\cite{feldmann2020lockdown}. In the United States, a blog
post~\cite{ncta2020covid} reveals that the national downstream peak traffic has
recently stabilized, but in the early weeks of the pandemic, it showed a growth of
20.1\%. For wireless networks in the US, volume increases of up to 12.2\% for voice
and 28.4\% for data by the top four providers were shown in an industry
report~\cite{ctia2020wireless}. Mobile networks in the UK reported roughly 25\%
drops in downlink data traffic volume~\cite{lutu2020characterization}. Industry
operators have self-reported on their network responses largely through blog
posts~\cite{comcast2020covid,att2020our,google2020keeping,mckeay2020parts}.

For traffic performance changes, different patterns appear in different
regions. Facebook shows that less-developed regions exhibited larger
performance degradations through their analysis of edge
networks~\cite{timm2020how}. Network latencies were approximately 30\% higher
during the lockdown in Italy~\cite{feldmann2020lockdown}.  According to an NCTA
report, networks in the United States saw less
congestion~\cite{ncta2020covid}. Due to decreased user mobility, cellular
network patterns have shifted~\cite{lutu2020characterization}: The authors
found a decrease in the average user throughput as well as decreased handoffs.
Feldmann et al.~\cite{feldmann2020lockdown} observed that the fixed-line
Internet infrastructure was able to sustain the 15--20\% increase in traffic
that happened rapidly during a short window of one week.

Our work differs from and builds on these previous studies in several ways:
First, this study extends over a longer time frame, and it also uses
longitudinal data to compare traffic patterns {\em during} the past six months
to traffic patterns in previous years.  Due to the nascent and evolving nature
of COVID-19 and corresponding ISP responses, previous studies have been
limited to relatively short time frames, and have mainly focused on Europe.
Second, this work explores the ISP {\em response} to the shifting demands and
traffic patterns; to our knowledge, this work is the first to begin to explore
ISP and service provider responses.

\paragraph{Application Measurements during COVID-19.}
Previous work has also studied application usage and performance, such as
increases in web conferencing traffic, VPN, gaming, and
messaging~\cite{feldmann2020lockdown}. Favale et al. studied ingress and egress
traffic from the perspective of a university network and found that the Internet
proved capable of coping with the sudden spike in demand in
Italy~\cite{favale2020campus}. Another paper used network traffic to determine
campus occupancy at the effect of COVID-19 related policies on three campus
populations across Singapore and the United States~\cite{zakaria2020analyzing}.
The cybercrime market was also statistically modeled during the COVID-19 era to
characterize its economic and social changes~\cite{vu2020turning}.

\paragraph{Network measurements of other disasters.}
While COVID-19 responses are ongoing and evolving, making measurement efforts 
incomplete, network responses under other disastrous events can be
informative. In 2011, the Japan earthquake of Magnitude 9.0 caused circuit
failures and subsequent repairs within a major ISP. Nationwide, traffic fell by
roughly 20\% immediately after the earthquake. However, surprisingly little
disruption was observed from outside~\cite{cho2011japan}. In 2012, Hurricane
Sandy hit the Eastern seaboard of the United States and caused regional outages
and variances over the network~\cite{heidemann2012preliminary}. For human-caused
disasters such as the September 11th attacks, routing, and protocol data were
analyzed to demonstrate the resilience of the Internet under stress. Their findings
showed that although unexpected blackouts did happen, they only had a local
effect~\cite{partridge2003internet}. Oppressive regimes have also caused
Internet outages, such as a complete Internet shutdown due to censorship actions
during the Egypt and Libya revolts~\cite{dainotti2011analysis}, where packet
drops and BGP route withdrawals were triggered intentionally.

Although there have been several preliminary measurements of the effects of
the COVID-19 response, none have holistically studied traffic data,
performance analysis, routing data, and ISP capacity information together, as
we do in this paper. It is crucial to collect and correlate such information
to better understand the nature of both traffic demands, the effects of these
changes on performance, and the corresponding responses.  This paper does so,
illuminating the collaborative view of responses of service providers in the
United States.

%% file: data.tex
\section{Data}\label{sec:data}

We leverage multiple network traffic datasets to facilitate our study:

\paragraph{Traffic Demands and Interconnect Capacity: Internet Connection Measurement Project.}
We leverage a dataset that includes network interconnection statistics for links
between anonymized access ISPs and their neighboring partner networks in the
United States. The dataset consists of IPFIX-based statistics over five-minute
intervals, including: timestamp, region (as access ISPs may connect to a partner
network in multiple geographic regions), anonymized partner network, access ISP,
ingress bytes, egress bytes, and link capacity. The dataset contains roughly
97\% of links from all participating ISPs. All of the links represented in the
dataset are private (i.e., they do not involve public IXP switch fabrics). Each
five-minute interval includes the sum of the utilization of traffic flows that were
active during that interval. We also calculate secondary statistics from the
dataset, including: timestamp for the peak ingress and egress hour for each day
on each link in terms of usage, ingress/egress peak hour bytes, and daily 95th
and 99th percentile usage.  

\paragraph{Performance Data: Federal Communications Commission Measuring Broadband America (MBA).} 
We analyze the FCC's ongoing nationwide performance measurement of broadband
service in the United States~\cite{MBA:2020}. The raw data is collected from a
collection of distributed measurement devices (named Whiteboxes) placed in
volunteer's homes and operated by SamKnows. Measurements are conducted on an
hourly basis. The dataset includes raw measurements of several performance
metrics, such as timestamp, unit ID, target server, round trip time, traffic
volume, etc. Each Whitebox also includes information pertaining to its ISP,
technology, and state where it is located. We also define dates related to the
status of the pandemic response (e.g., stay-at-home orders, state of
emergency declaration, etc.). Based on these, we can compute more
statistics for specified groups (e.g., break into ISPs): average and standard
deviation among Whiteboxes, daily 95th and 99th percentile latency/throughput. 

\begin{figure}[t!]
    \centering
    \subfloat[Absolute utilization.\label{fig:raw_usage}]{%
      \includegraphics[width=0.9\columnwidth]{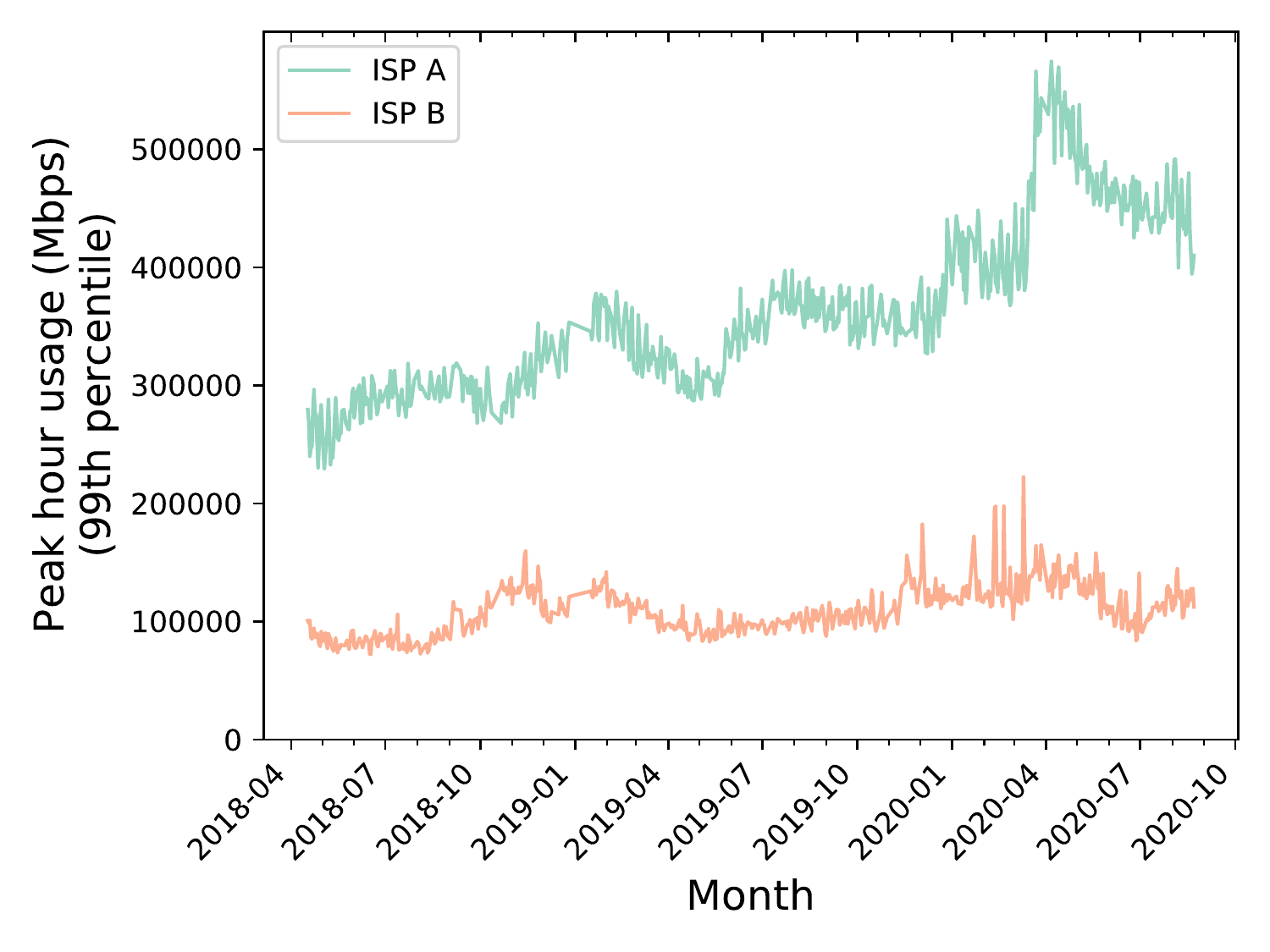}%
    }
    \hfill
    \subfloat[Normalized utilization.\label{fig:normal_usage}]{%
      \includegraphics[width=0.9\columnwidth]{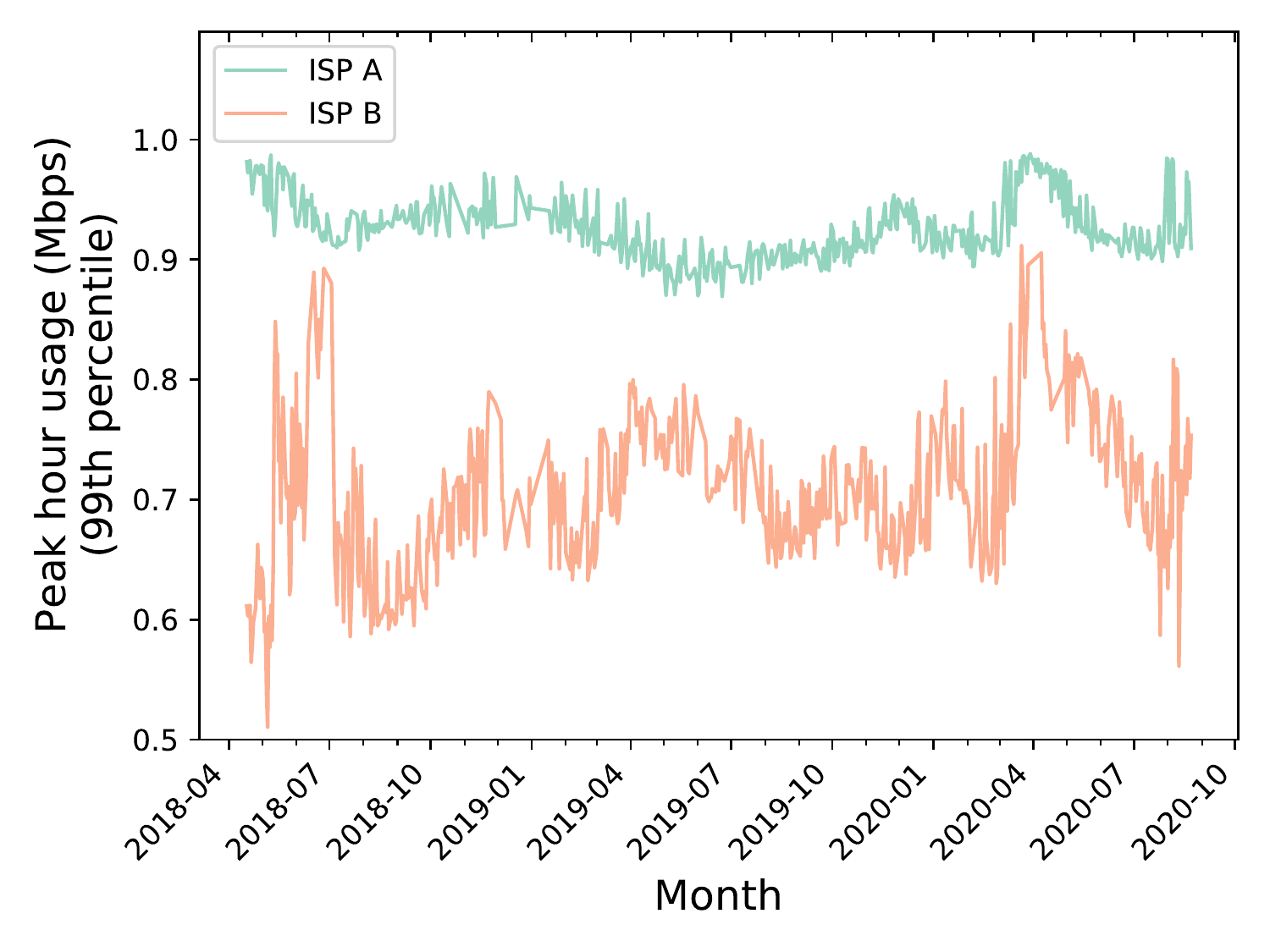}%
    }
    \caption{99th percentile interconnect link utilization for two ISPs.}
    \label{fig:usage}    
  \end{figure}

To keep the network capacity consistent and to record eventual changes solely
based on utilization factors, we pre-process the MBA dataset with several
filters. First, we filter the non-continuous data within the dates of interest
(Dec. 1st, 2019 to June, 30th 2020, and the previous year) to capture 
successive shifts. Then, we eliminate the Whiteboxes 
which do not aggregate a statistically significant amount of data, such as some 
states, ISPs, and technologies with limited data (e.g., satellite).
Finally, we choose the measurements from Whiteboxes to the top 10 most targeted 
servers across the United States to represent the overall US performance. These 
servers are sparsely located in major cities of the US and they have the most 
Whiteboxes (over 200 for each ISP) connecting with them.

\paragraph{IP Prefix Advertisements: RouteViews.} 
To gain insight into changes in IP address space, we parse Internet-wide BGP
information globally from several locations and backbones via RouteViews. Raw
RIBs (Routing Information Bases) files were obtained from
RouteViews~\cite{route_views:2020} data on a weekly basis. The average of each
Tuesday is computed to represent that week. The RIBs are then parsed to obtain
IPv4 Prefix-to-Autonomous System (AS) relationships, including mappings of IP
prefix, prefix length, paths of AS numbers. In Section~\ref{sec:ip_space}, we
compute the total advertised IPv4 spaces for AS numbers associated with two
popular video conferencing applications: Zoom and Cisco
WebEx~\cite{fcc2020keep}.

%% file: traffic.tex
\section{How did traffic demands change?}\label{sec:traffic}

Because most previous
studies~\cite{feldmann2020lockdown,lutu2020characterization,candela2020impact}
focus on Europe, we begin our explorations by validating whether
similar traffic changes are observed in the United States. We consider peak hour
link utilization from the Interconnect Measurement Project as a measure of
traffic demand. We pre-process the interconnect dataset and remove
anomalous data points that are caused by failures in the measurement system. In
particular, we do not analyze dates that are greater than two standard
deviations outside of a 60-day rolling mean for each link. 
  
Figure~\ref{fig:usage} shows both the absolute utilization and the utilization
normalized against the link capacity for two anonymized ISPs. For each ISP, we
plot the value corresponding to the 99th percentile link utilization for a given
day. We observe from Figure~\ref{fig:raw_usage} that ISP A saw a dramatic
increase in raw utilization at roughly the same time as the initial COVID-19
lockdowns (early March 2020), with values tapering off slightly over the summer
of 2020. ISP B, on the other hand, saw a smaller raw increase in utilization for
its 99th percentile links. To better understand whether ISP B's smaller increase
is a byproduct of different operating behaviors, we explore possible trends in
the normalized data (Figure~\ref{fig:normal_usage}). Here we see that both ISPs
experienced significant increases in utilization in March and April 2020.  

\begin{figure}[t!]
  \centering
  \subfloat[Peak download.\label{fig:peakdw_peer}]{%
    \includegraphics[trim=0cm 0cm 0cm 1.8cm, clip=true, width=0.8\columnwidth]{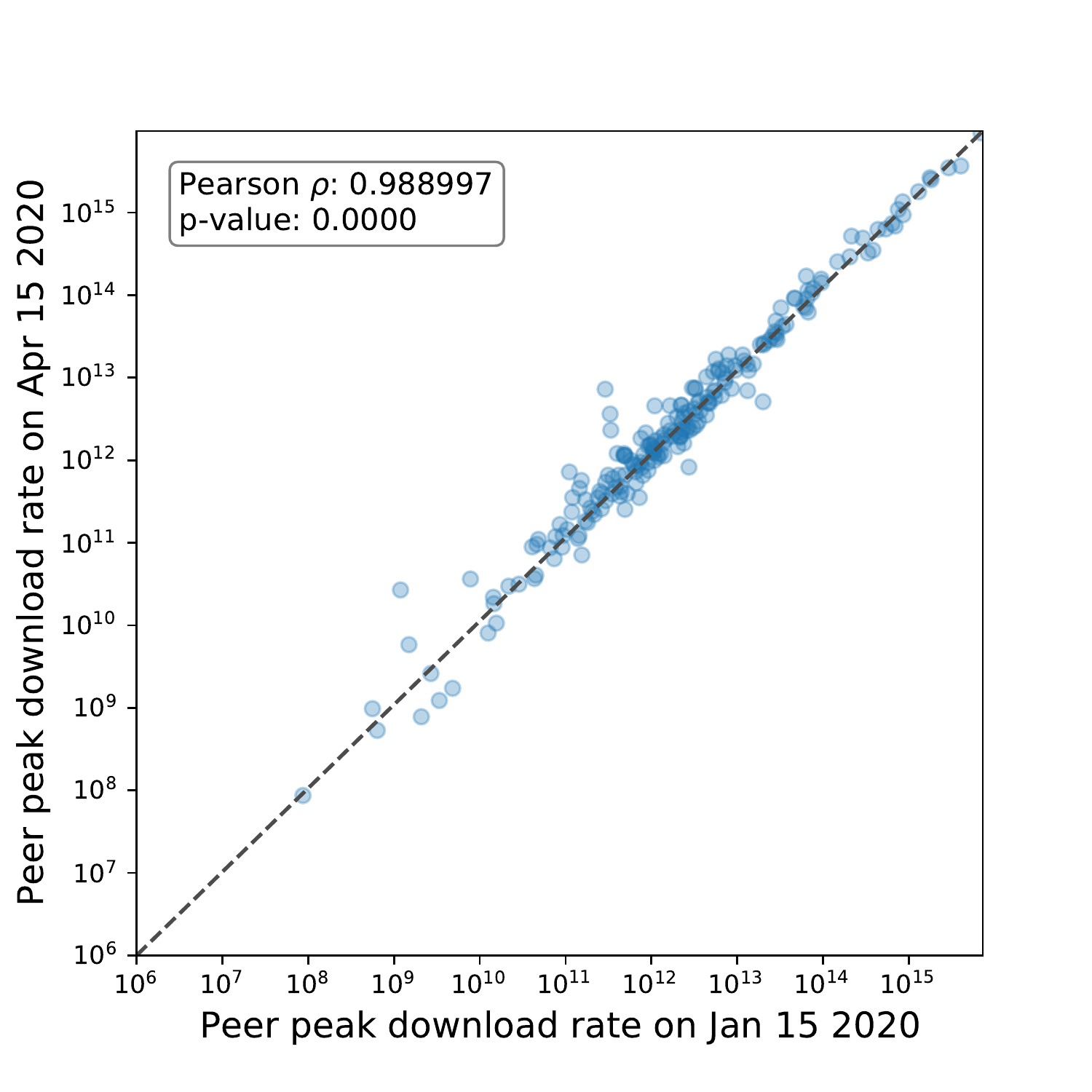}%
  }
  
  \subfloat[Peak upload.\label{fig:peakup_peer}]{%
      \includegraphics[trim=0cm 0cm 0cm 1.8cm, clip=true, width=0.8\columnwidth]{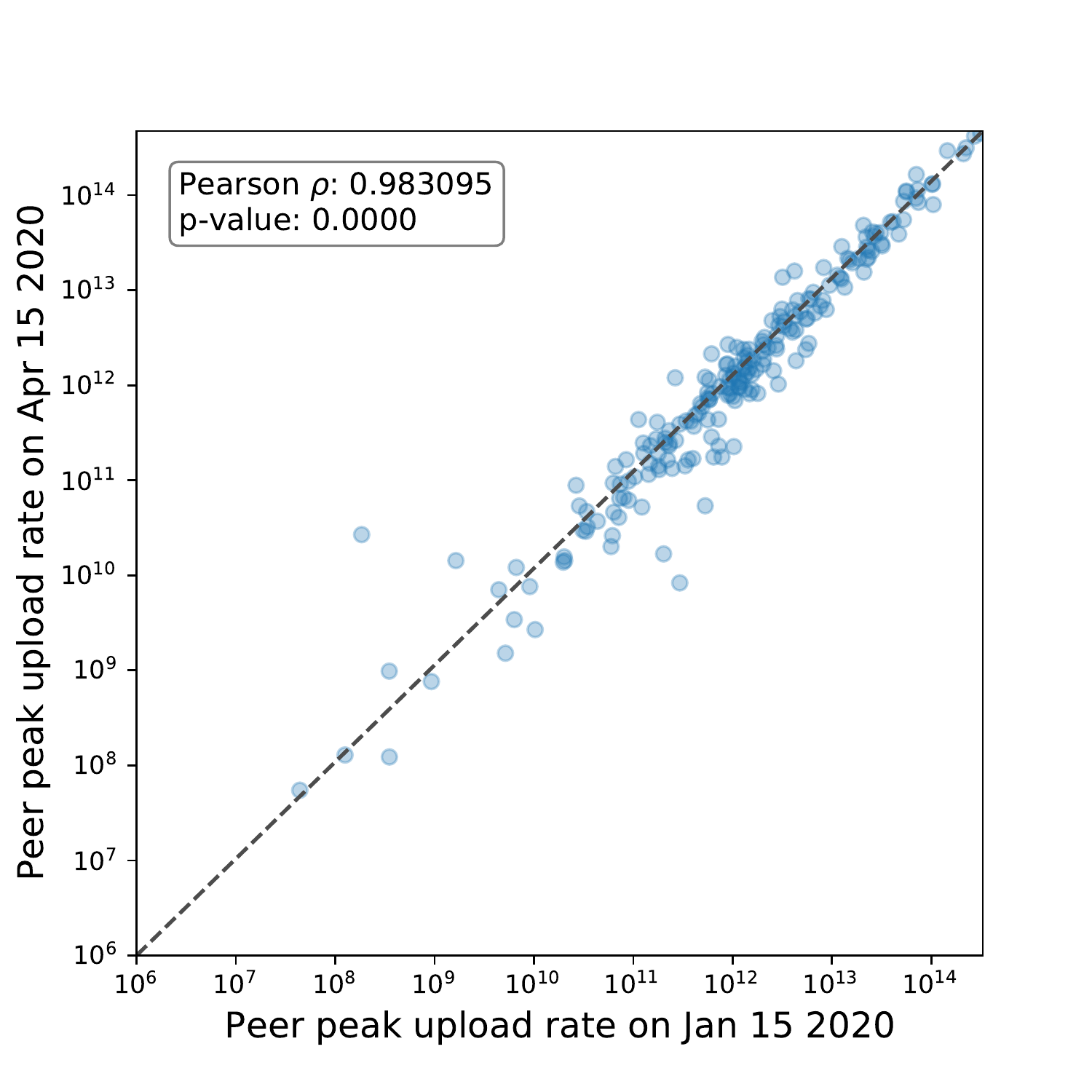}%
  }
  
  \caption{Peer link utilization for ISP A between January 15 to April 15, 2020.}
  \label{fig:peak}    
  \end{figure}

We also investigated how traffic patterns changed between ISP A and each of
its peers, in both the upstream and downstream directions.
For this analysis, we focused on the dates around the utilization
peaks shown in Figure~\ref{fig:usage}. We compared the peak hour download and
upload rates on all of ISP A's interconnects on (1)~January 15, 2020, and (2)~April
15, 2020 (Figure~\ref{fig:peak}). In general, we see that traffic patterns to
peers do not vary greatly between the two dates. We do see, however, that
traffic volumes to (and from) some peers 
change significantly---some by several orders of
magnitude. The identities of the peers are anonymous in the dataset, 
but some patterns are nonetheless clear: For example, some peers show an
increase of upstream utilization by two or three orders of magnitude.
Such drastic changes may be attributable to users working from home
and connecting to services that would cause more traffic to traverse the peer
link in the upstream direction.
We confirmed these results with the operators at ISP A and report that they
observed that streaming video traffic decreased from 67 to 63\% of the total
traffic, but video conferencing increased from 1\% to 4\% as a percentage of
overall traffic.

%% file: performance.tex
\section{What was the effect on performance?}\label{sec:performance}

The surge in interconnect utilization poses a challenge for service providers,
as high utilization of interconnects can potentially introduce high delays for
interactive traffic, packet loss, or both.  These effects can ultimately be
observed through changes in latency (and, potentially, short-term throughput).
To examine whether we can observe these effects, we look into the latency and
throughput reported by the Measuring Broadband America (MBA)
dataset~\cite{MBA:2020}. We explore these effects over the course of several
years to understand whether (and how) performance anomalies that we observe
during COVID-19 lockdown differ significantly from performance anomalies
observed during other time periods.

\subsection{How performance changed after lockdown.}
To better understand how performance changed as a result of changing traffic
patterns during the COVID-19 lockdown in the United States, we explored how
latency evolved over the course of 2020.  To establish a basis for comparison,
we show the time period from late 2019 through mid-2020. The Appendix also
contains a similar analysis for the 2018--2019 time period. We compute the
average latency per-Whitebox per-day, and subsequently explore distributions
across Whiteboxes for each ISP. (As discussed in Section~\ref{sec:data}, we
consider only Whiteboxes in fixed-line ISPs for which there are an adequate
number of Whiteboxes and samples.) We use March 10th\footnote{note that this 
is also the launch date of Call of Duty Warzone}, the average declaration 
of emergency date~\cite{wiki_dates:2020}, to mark the beginning of the COVID-19
pandmic phase (red shaded for figures).

\subsubsection{Longitudinal evolution of aggregate, average round-trip latency.}  

\begin{figure*}[t]
  \centering
    \includegraphics[width=0.7\textwidth]{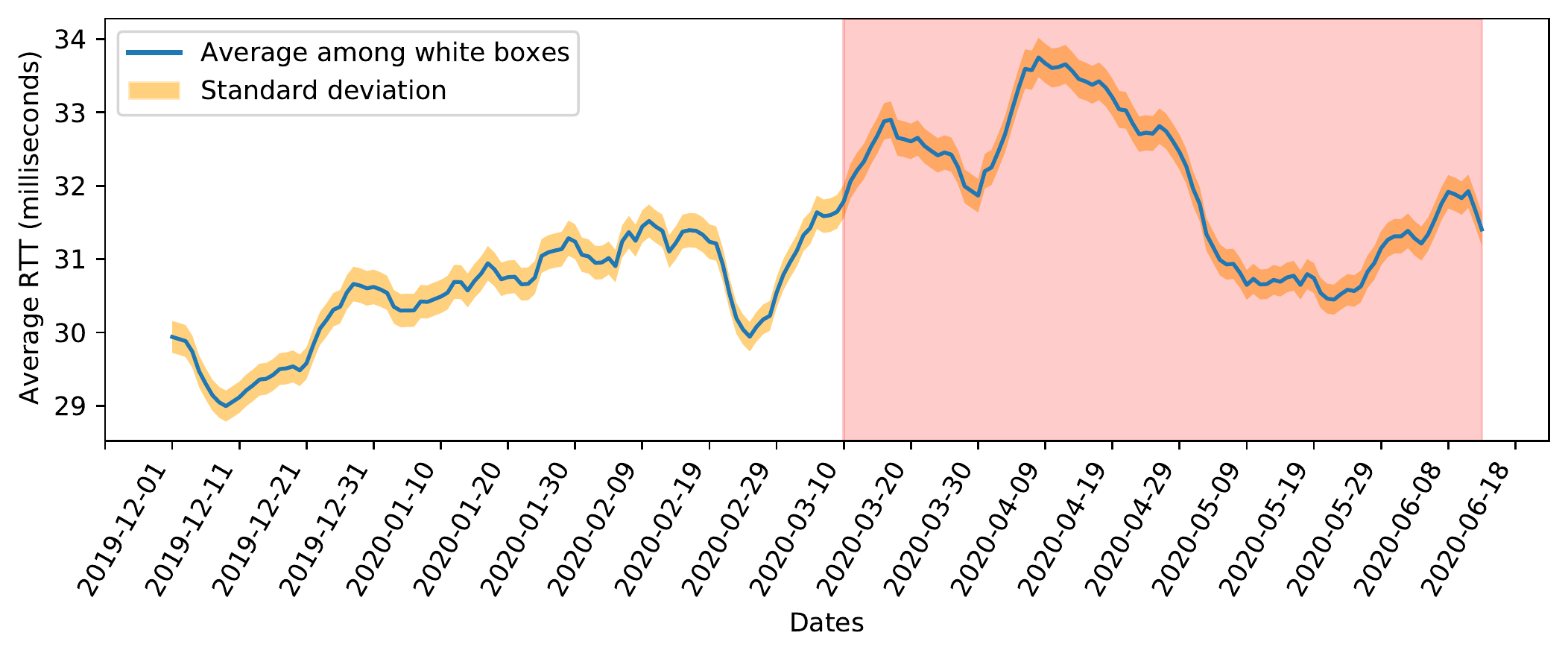}
    \caption{Daily changes of latency from Dec. 2019 to June 2020.
    The pandemic phase is marked in red. Change in average latency across the
    non-satellite ISPs in the FCC MBA program reflect a small (2--3~ms) but
    significant increase in overall average latency. (Note: y-axis does not start at zero.)
    }
    \label{fig:daily}
\end{figure*}

Figure~\ref{fig:daily} shows a seven-day moving average of average round-trip
latencies between all Whiteboxes in this study.
We observe an increase in average round-trip latency by as much as 10\%, this
increase in mean latency is significant,
corresponding to 30x standard deviation among all Whiteboxes. 
At the end of April, latencies return to early 2020 levels. It is worth noting
that, although this increase in average latency is both sizable and
significant, similar deviations and increases in latency have been observed
before (see the Appendix for comparable data from 2018--2019). Thus, although
some performance effects are visible during the COVID-19 lockdown, the event
and its effect on network performance are not significantly different from
other performance aberrations. Part of the reason for this, we believe, may be
the providers' rapid response to adding capacity during the first quarter of
2020, which we explore in more detail in Section~\ref{sec:capacity}.

\subsubsection{Longitudinal evolution of per-ISP latencies.} 
\begin{figure*}[t]
  \centering
  \subfloat[95th percentile of ISP latency (Group 1).\label{fig:ISP_G1_95}]{%
    \includegraphics[width=0.7\textwidth]{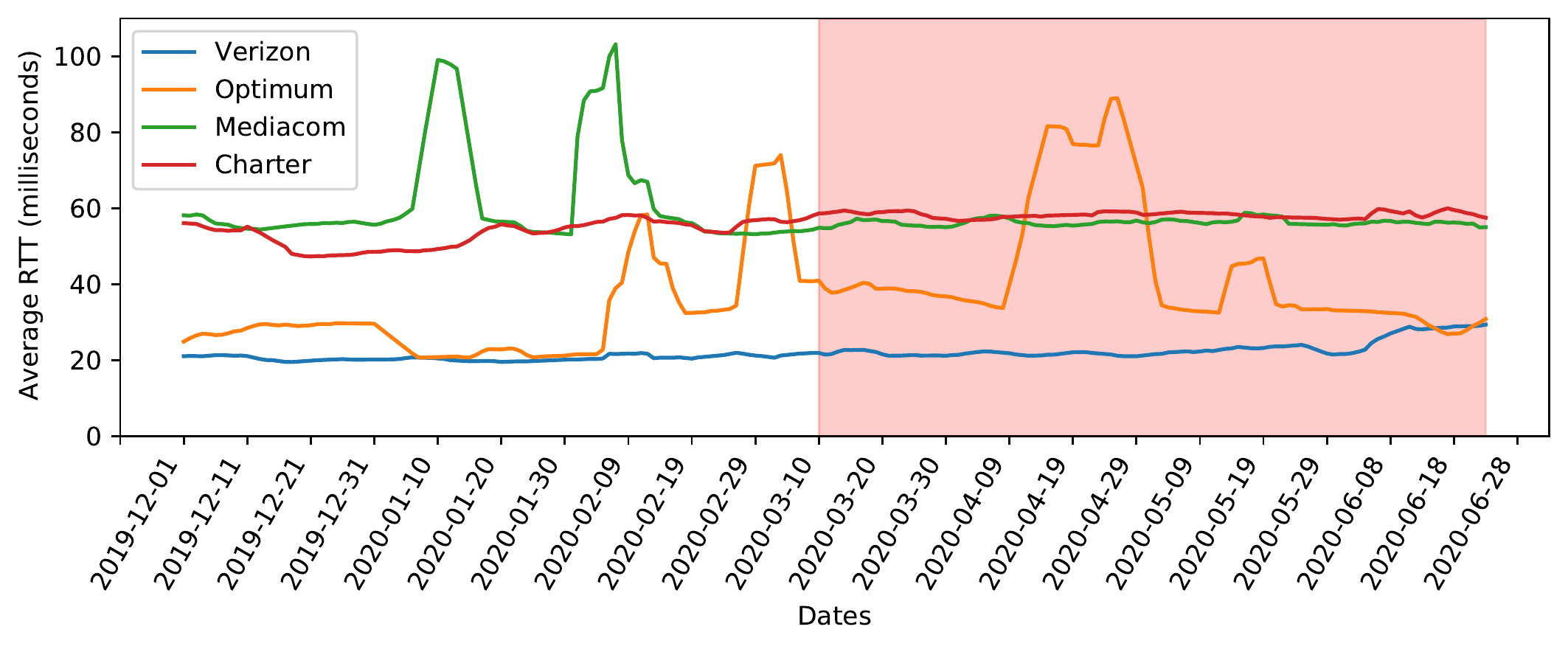}%
  }\hfil
  
  \subfloat[95th percentile of ISP latency (Group 2).\label{fig:ISP_G2_95}]{%
    \includegraphics[width=0.7\textwidth]{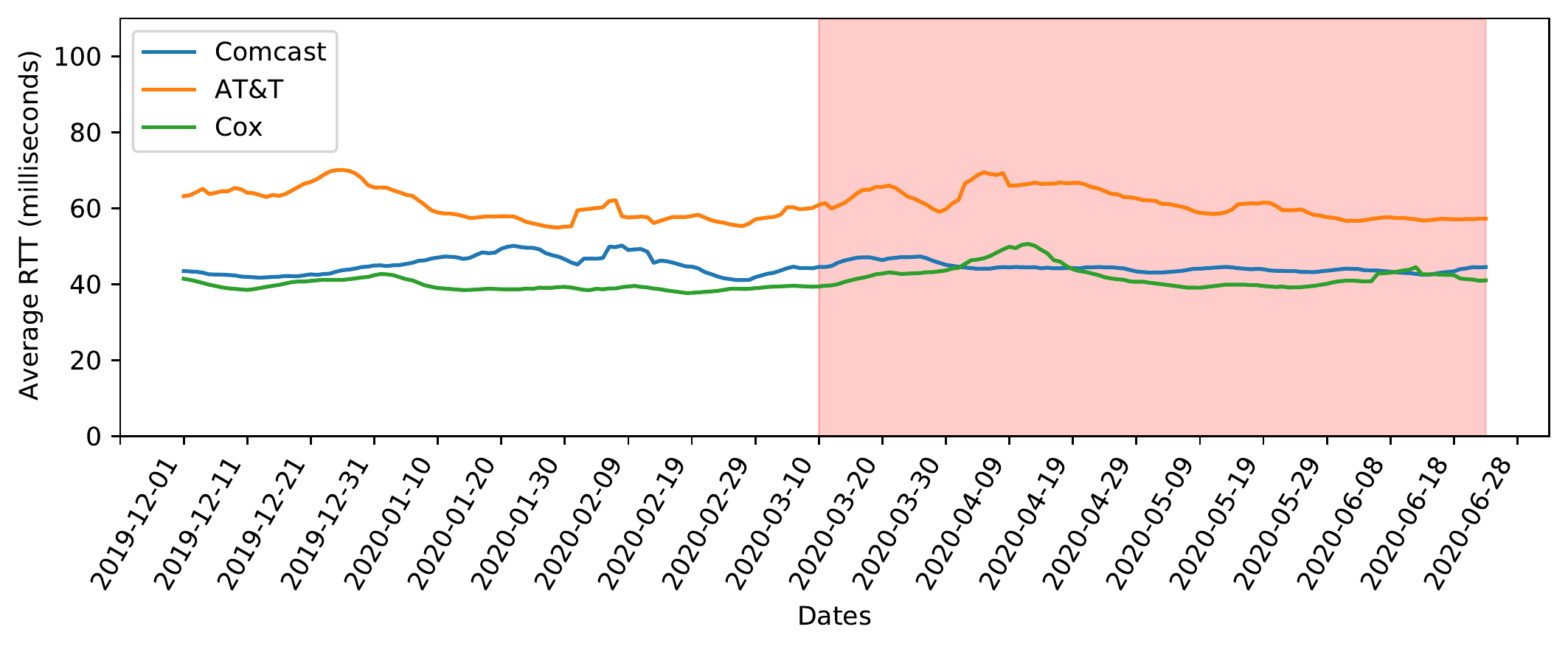}%
  }\hfil
  
  \caption{Latency (95th percentile) for different ISPs.}
  \label{fig:ISP_latency_95}
  
\end{figure*}

\begin{figure*}[t]
  \centering
  \subfloat[99th percentile of ISP latency (Group 1).\label{fig:ISP_G1_99}]{%
    \includegraphics[width=0.7\textwidth]{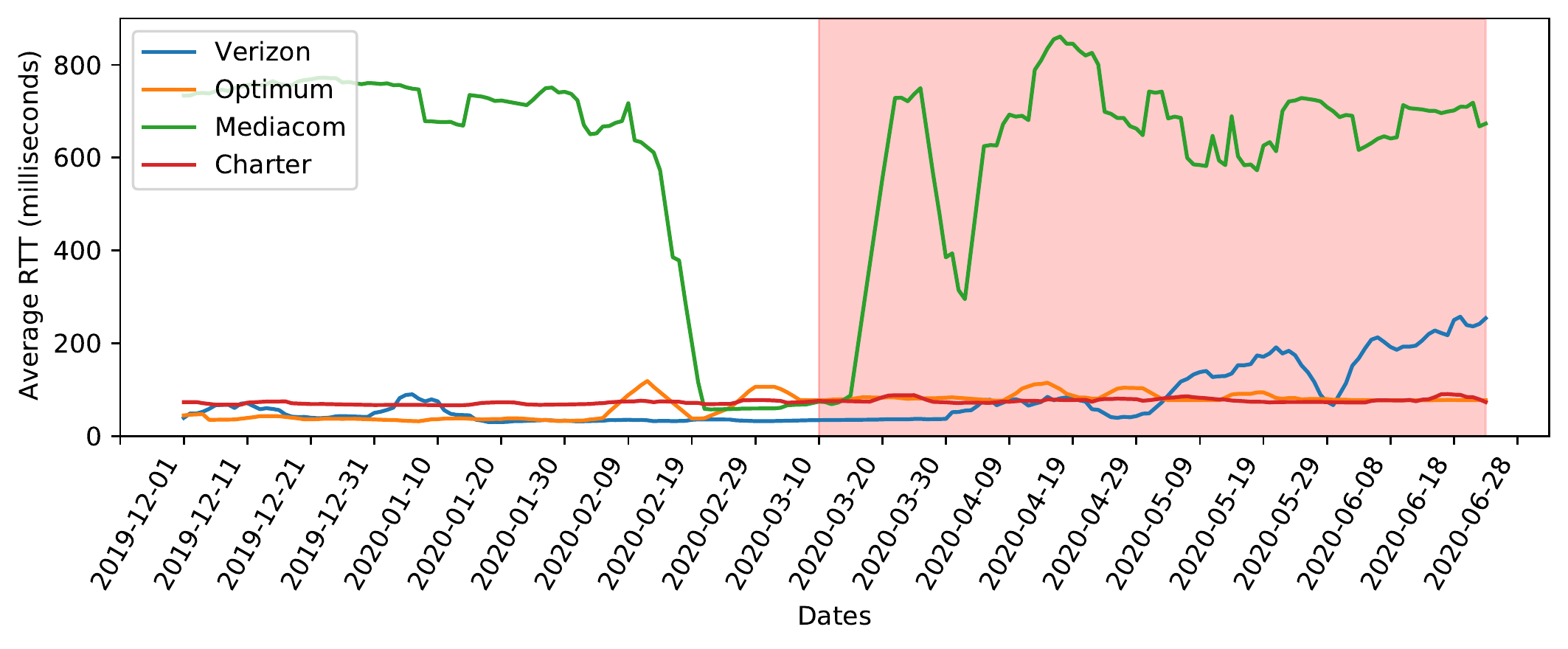}%
  }\hfil
  
  \subfloat[99th percentile of ISP latency (Group 2).\label{fig:ISP_G2_99}]{%
    \includegraphics[width=0.7\textwidth]{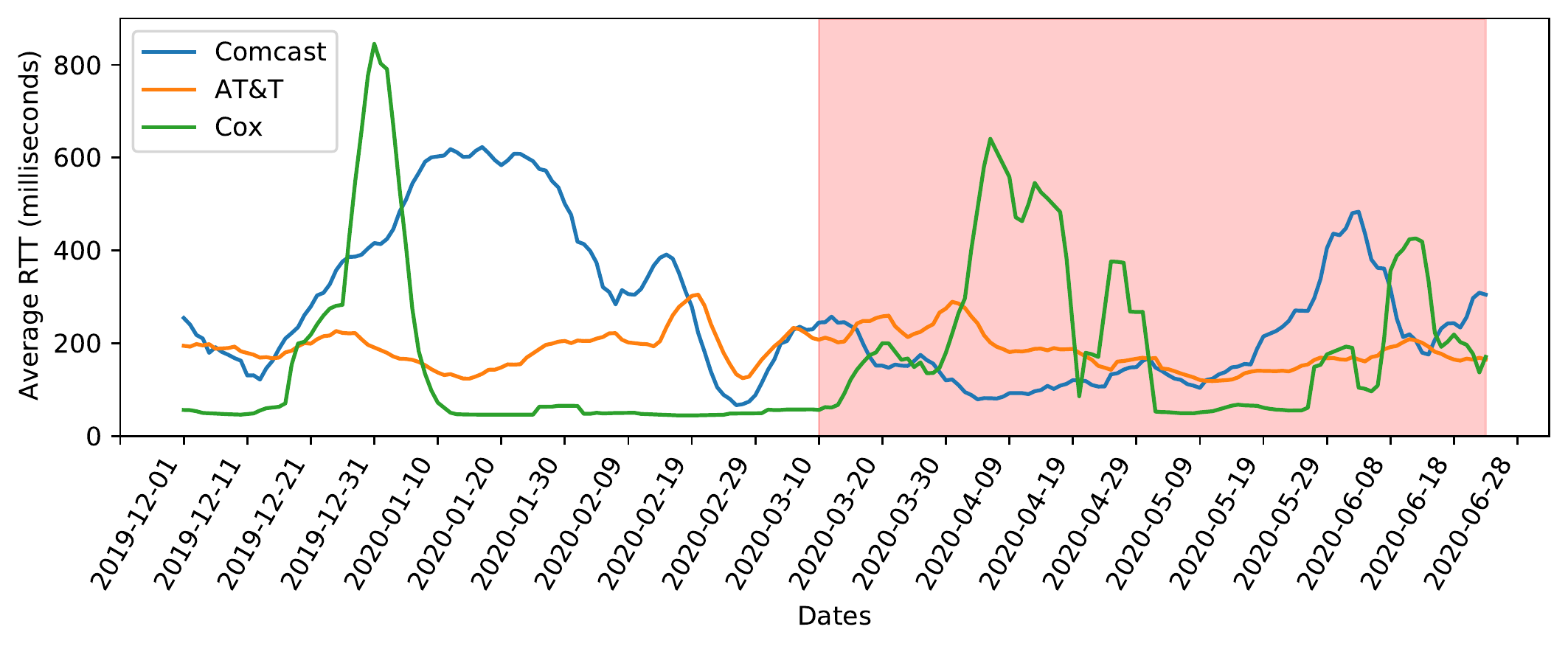}%
  }\hfil
  
  \caption{Latency (99th percentile) for different ISPs.}
  \label{fig:ISP_latency_99}
  
\end{figure*}

In addition to the overall changes in performance, we also explored per-ISP
latency and throughput effects before and during the COVID-19 lockdown period.
Figures~\ref{fig:ISP_latency_95} and~\ref{fig:ISP_latency_99} show these
effects, showing (respectively) the 95th and 99th percentiles of average round-trip latency
across the Whiteboxes. These results show that, overall 95th percentile
latency across most ISPs remained stable; 99th percentile latency, on the
other hand, did show some deviations from normal levels during lockdown for
certain ISPs. Notably, however, in many cases the same ISPs experienced
deviations in latency during other periods of time, as well (e.g., during the
December holidays).

\subsection{Throughput-latency relationship}

High latencies can sometimes be reflected in achieved throughput, given the
inverse relationship between TCP throughput and round-trip latency.  To
explore whether latency aberrations ultimately result in throughput effects,
as well as how those effects manifest at different times of day, we explored
the distribution of latencies before COVID-19 emergency declarations (ED),
after the ED but before the stay-at-home order (SO). Our hypothesis was that
we might see higher latencies (and lower throughputs) during ``peak hours'' of
the day from broadband access networks, with the peak hours effectively
expanded to the weekday working hours, in accordance with previous
descriptions of these effects~\cite{comcast2020covid}.

We explored these metrics for a baseline period predating COVID-19, the time
between state declaration of emergency and stay at home
ordered~\cite{wiki_dates:2020}, after stay-at-home declarations were ordered,
and two months after stay-at-home ordered. Because these dates differed across
states, we used known dates for each state~\cite{wiki_dates:2020} and matched
the corresponding dates for each state against the known location of the
Whiteboxes.

Figure~\ref{fig:time} shows the distribution throughput and latency
distributions across all Whiteboxes for
four time intervals, 
plotted in four-hour intervals. From Figure~\ref{fig:hourly_L}, it is clear that
the quantiles, median, and maximum latencies
all exhibit effects that correlate with these time periods.

The period between
ED and SO corresponds to abrupt routing changes, and the latency data thus
reflects a corresponding degradation during this time interval, perhaps at
least partially due to the fact that providers cannot immediately respond
after the initial emergency declaration (we discuss the timeframes during
which capacity was added to the networks in Section~\ref{sec:capacity}).
As the transition continues, SO appears to be a point in time where latency
stabilizes.
Figure~\ref{fig:hourly_T} shows that distributions of throughput measurements
are more robust, although the upper end of the distribution is clearly
affected, with maximum achieved throughputs lower.
The median and minimum have negligible changes during time periods in late 
April suggesting (and corresponding to) aggressive capacity augmentation,
which we discuss in more detail in Section~\ref{sec:capacity}.

%% file: capacity.tex
\section{How did service providers respond?}\label{sec:capacity}

In this section, we study how service providers responded to the changes in
traffic demands. We focus on the capacity changes during lockdown by inspecting
two data sources: (1)~to understand how ISPs responded by adding capacity to
interconnects, we study the interconnect capacity of two large ISPs in the
United States; and (2)~to understand how video service providers expanded their
network footprints in response to increasing demand, we analyze IPv4 address
space from two major video conference providers---WebEx and Zoom---and find that
both providers substantially increased advertised IP address
space.

\subsection{Capacity increases at interconnect} 

\begin{figure}[h!tp]
    \centering
    \subfloat[ISP A\label{fig:comcast_cap}]{%
      \includegraphics[width=0.9\columnwidth]{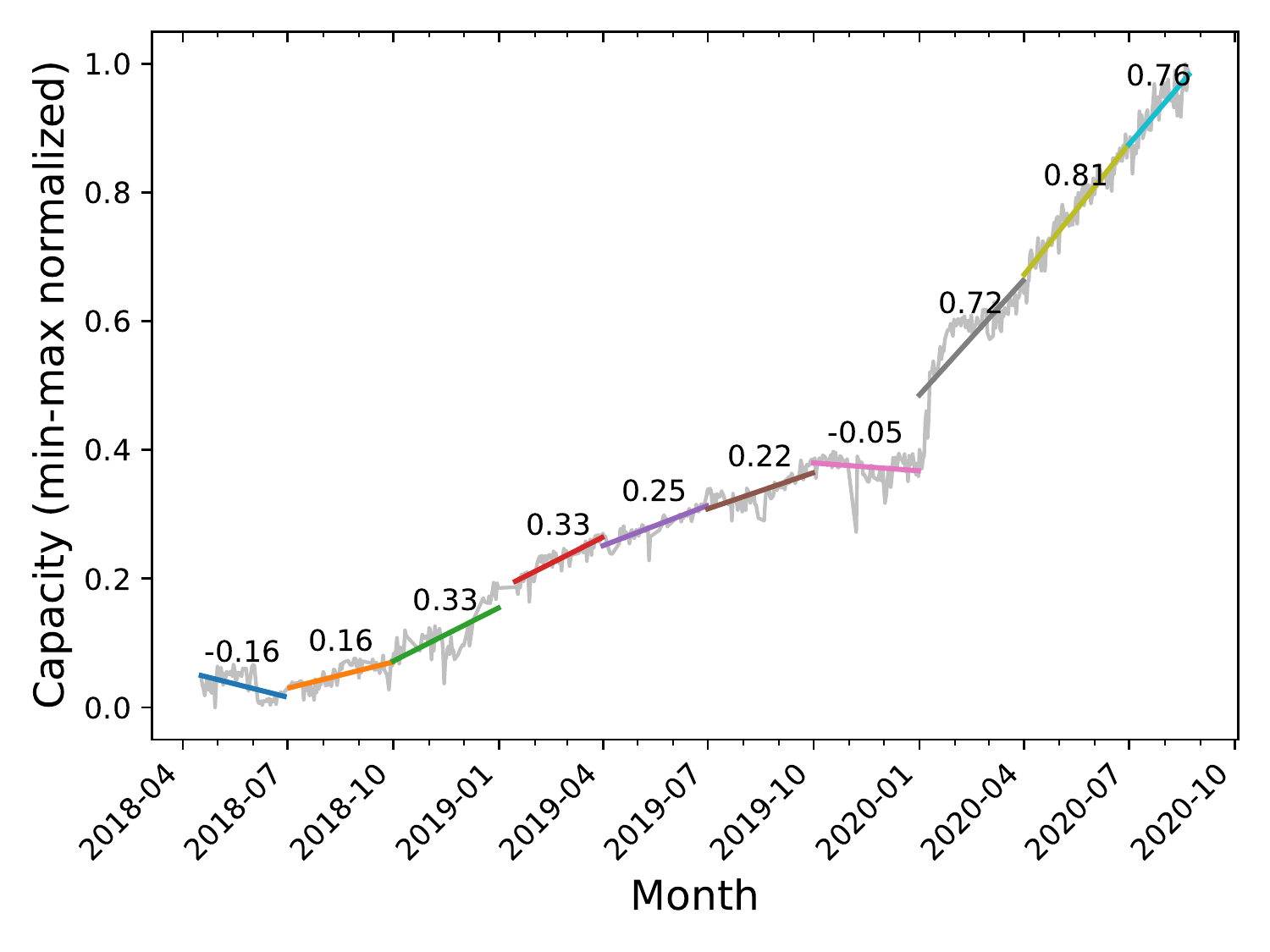}%
    }\hfill
    \subfloat[ISP B\label{fig:mediacom_cap}]{%
      \includegraphics[width=0.9\columnwidth]{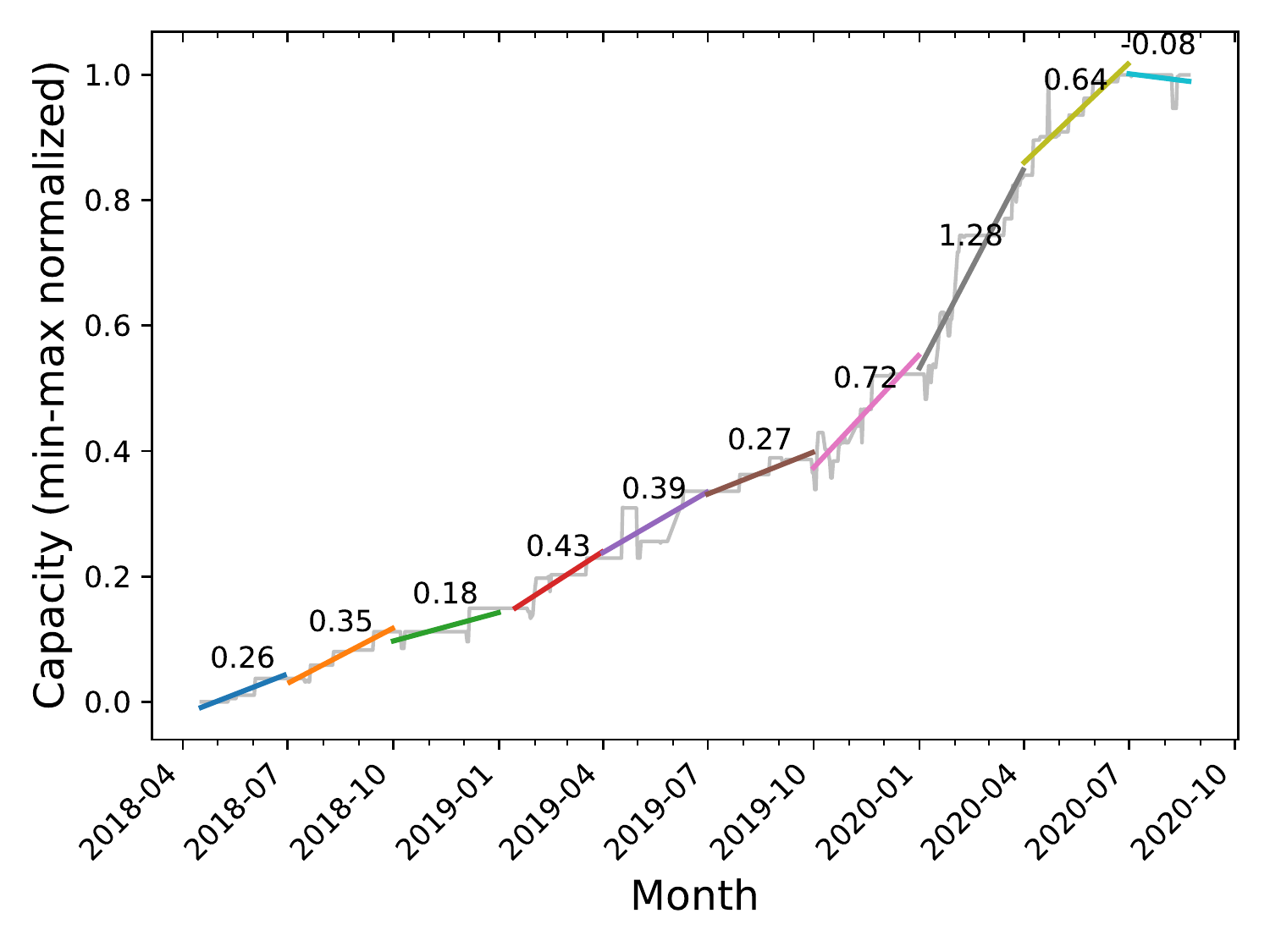}%
    }
    
    \caption{Normalized interconnect capacity increases for two ISPs.}
    \label{fig:cap}    
\end{figure}

We begin by exploring how ISPs responded to changing traffic demands by adding
network capacity at interconnect links. To do so, we use the Interconnect
Measurement Project dataset. We calculate the total interconnect capacity for
each ISP by summing the capacities for all of the links associated with the ISP.
To enable comparison between ISPs that may have more or less infrastructure
overall, we normalize the capacity values for each using min-max normalization.
We again filter out date values that are beyond two standard deviations from a
rolling 60-day window mean. To show aggregate infrastructure changes over time,
we take all of the data points in each fiscal quarter and perform a
least-squares linear regression using SciKit Learn. This regression yields a slope
for each quarter that illustrates the best-fit rate of capacity increases over
that quarter. We scale the slope value to show what the increase would be if the
pace was maintained for 365 days (i.e., a slope of 1 would result in a doubling
of capacity over the course of a year). Figure~\ref{fig:cap} shows the resulting
capacity plots.

The overall trend shows how these two ISPs in the United States aggressively
added capacity at interconnects---at more than twice the rate at which they were
adding capacity over a comparable time period in the previous year. Second, both
ISPs significantly added capacity in the first quarter of 2020---at a far
greater rate than they were adding capacity in the first quarter of 2019. Recall
from the usage patterns shown in Figure~\ref{fig:usage}, ISP A tends to
operate their links at nearly full capacity, in contrast to ISP B, where
aggregate utilization is well below 90\%. Both ISPs witnessed a jump in usage
around the lockdown; the response of aggressively adding capacity appears to
have mitigated possible adverse effects of high utilization rates. The increase
in capacity was necessary to cope with the increased volume: although network
performance and utilization ratios returned to pre-COVID-19 levels, 
the {\em absolute} traffic volumes remain high.

\subsection{Increased advertised IP address space}\label{sec:ip_space}

To cope with abrupt changes caused by COVID-19, application service providers
also took action to expand their infrastructure. Previous work has observed
shifted traffic in communication applications (such as video conferencing apps,
email, and messaging) after lockdown~\cite{feldmann2020lockdown}. It has been
reported informally that many application providers expanded serving
infrastructure, changed the routes of certain application traffic flows, and
even altered the bitrates of services to cope with increased utilization.

While not all of these purported responses are directly observable in public
datasets; however, RouteViews makes available global routing information, which
can provide some hints about routes and infrastructure, and how various
characteristics of the Internet routing infrastructure change over time. This
data can provide some indication of expanding infrastructure, such as the amount
of IPv4 address space that a particular Autonomous System (AS) is advertising.
In the case of video conference providers, where some of the services may be
hosted on cloud service providers or where the video service is a part of a
larger AS that offers other services (e.g., Google Meet), such a metric is
clearly imperfect, but it can offer some indication of response.

To understand how service providers announced additional IPv4 address space, we
parsed BGP routing tables from
RouteViews~\cite{route_views:2020}. For each route that originates from ASes of
certain application providers, we aggregate IP prefixes and translate the
resulting prefixes into a single count of overall IPv4 address space.  
We focus on two popular video conferencing applications, Zoom and WebEx, since
they are two of the largest web conference providers in the United States---as
also recognized by the FCC in their recent order for regulatory
relief~\cite{fcc2020keep}. We track the evolution of the advertised IP address space
from the beginning of 2019 through October 2020.

\begin{figure}[t]
  \centering
    \includegraphics[width=0.9\columnwidth]{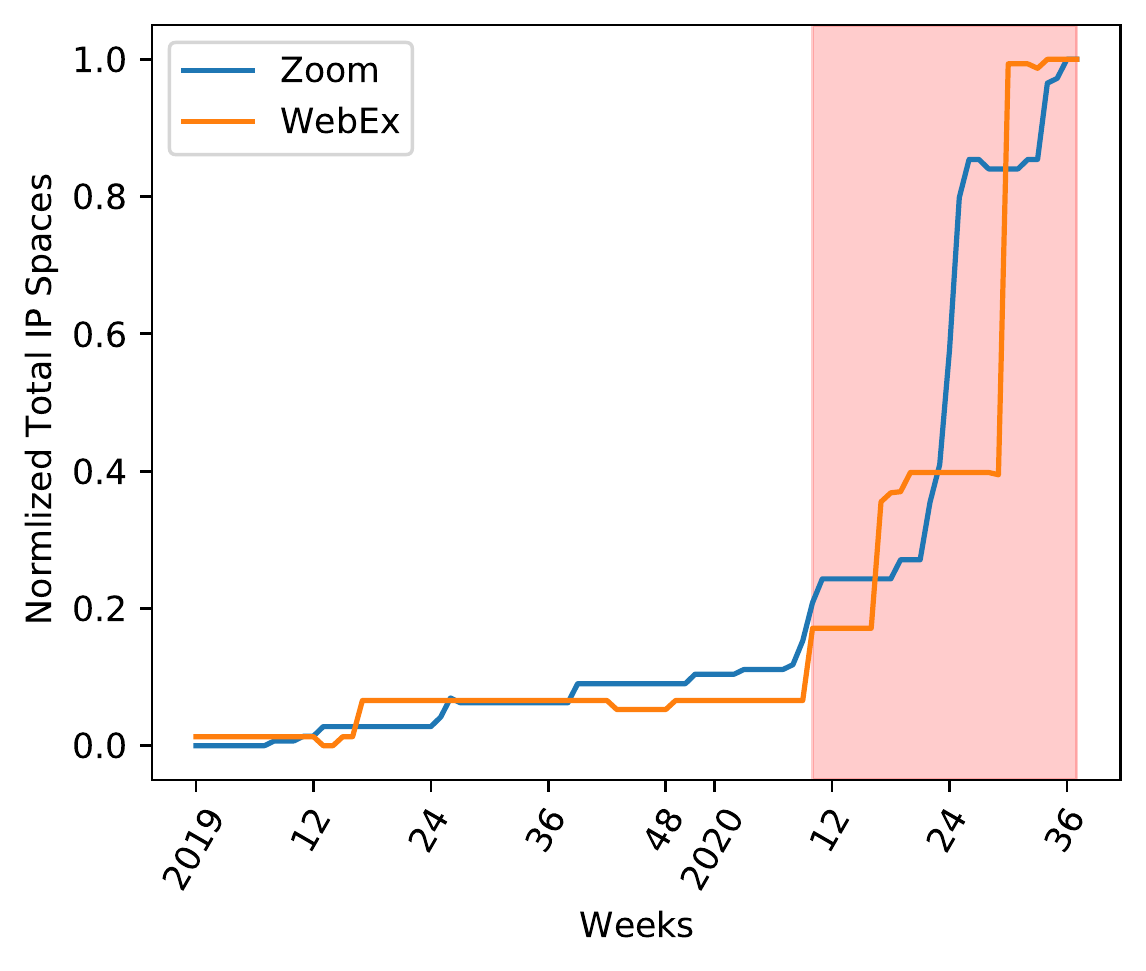}%
    \caption{Normalized advertised IPv4 space. 
    Red: COVID-19 pandemic phase.}\label{fig:IPSpace}
\end{figure}

\begin{table}[h!]
  \caption{Advertised IPv4 space.}
  \begin{tabular}[b]{lrr}
    \toprule
    App & Min & Max \\ 
    \midrule
    Zoom & 9,472 & 46,336 \\ 
    WebEx & 110,080 & 265,728 \\ 
    \bottomrule
  \end{tabular}
  \label{tab:IPSpace}
\end{table}

Figure~\ref{fig:IPSpace} demonstrates how each provider added advertised IPv4
address space from before the pandemic through October 2020. After the beginning
of the COVID-19 pandemic, both Zoom and WebEx rapidly begin to advertise
additional IPv4 address space. Table~\ref{fig:IPSpace} enumerates the absolute
values of advertised IP address space: Zoom and WebEx increased advertised IP
address space by about 4x and 2.5x respectively, roughly corresponding to the
2--3x increase in video conferencing traffic.

%% file: conclusion.tex
\section{Conclusion}\label{sec:conclusion}

This paper has explored how traffic demands changed as a result of the abrupt
daily patterns caused by the COVID-19 lockdown, how these changing traffic
patterns affected the performance of ISPs in the United States, both in aggregate
and for specific ISPs, and how service providers responded to these shifts in
demand.  We observed a 30--60\% increase in peak traffic rates for two major
ISPs in the US corresponding with significant increases in latency in early weeks
of lockdown, followed by a return to pre-lockdown levels, corresponding with
aggressive capacity augmentation at ISP interconnects and the addition of IPv4
address space from video conferencing providers. Although this paper presented
the first known study of interconnect utilization and service provider
responses to changes in patterns resulting from the COVID-19 pandemic, this
study still offers a somewhat limited viewpoint into these effects and
characteristics. Future work could potentially confirm or extend these
findings by exploring these trends for other ISPs, over the continued lockdown
period, and for other service providers.

%% file: appendix.tex


    \renewcommand{\thesection}{\Alph{section}}%
    \renewcommand{\thepage}{A-\arabic{page}}
    \setlength{\parskip}{0.1em}

    \begin{figure*}[b]
        \centering
        \includegraphics[width=0.7\textwidth]{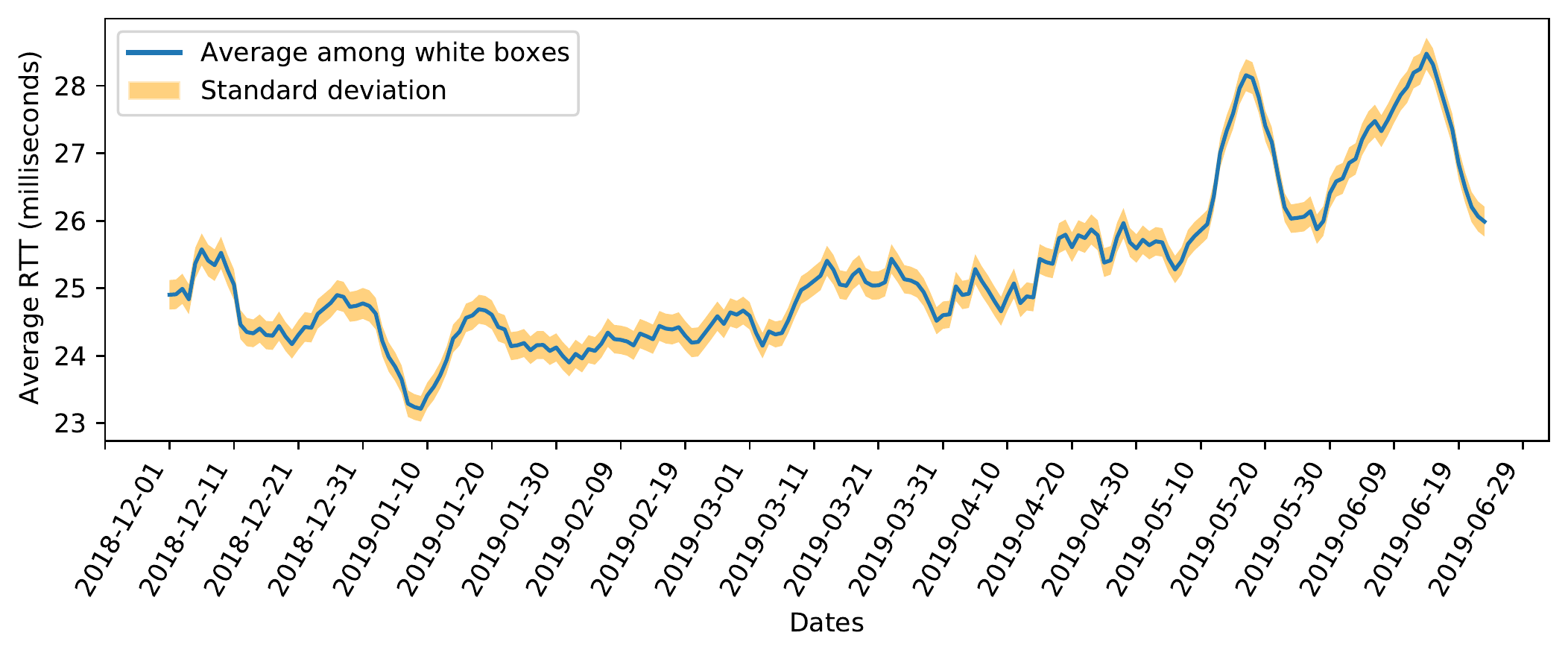}
        \caption{Daily changes of latency from Dec. 2018 to June 2019.
        (Note: y-axis does not start at zero.)
        }
        \label{fig:daily2}
        \subfloat[Latency.\label{fig:hourly_L}]{%
      \includegraphics[width=0.5\textwidth]{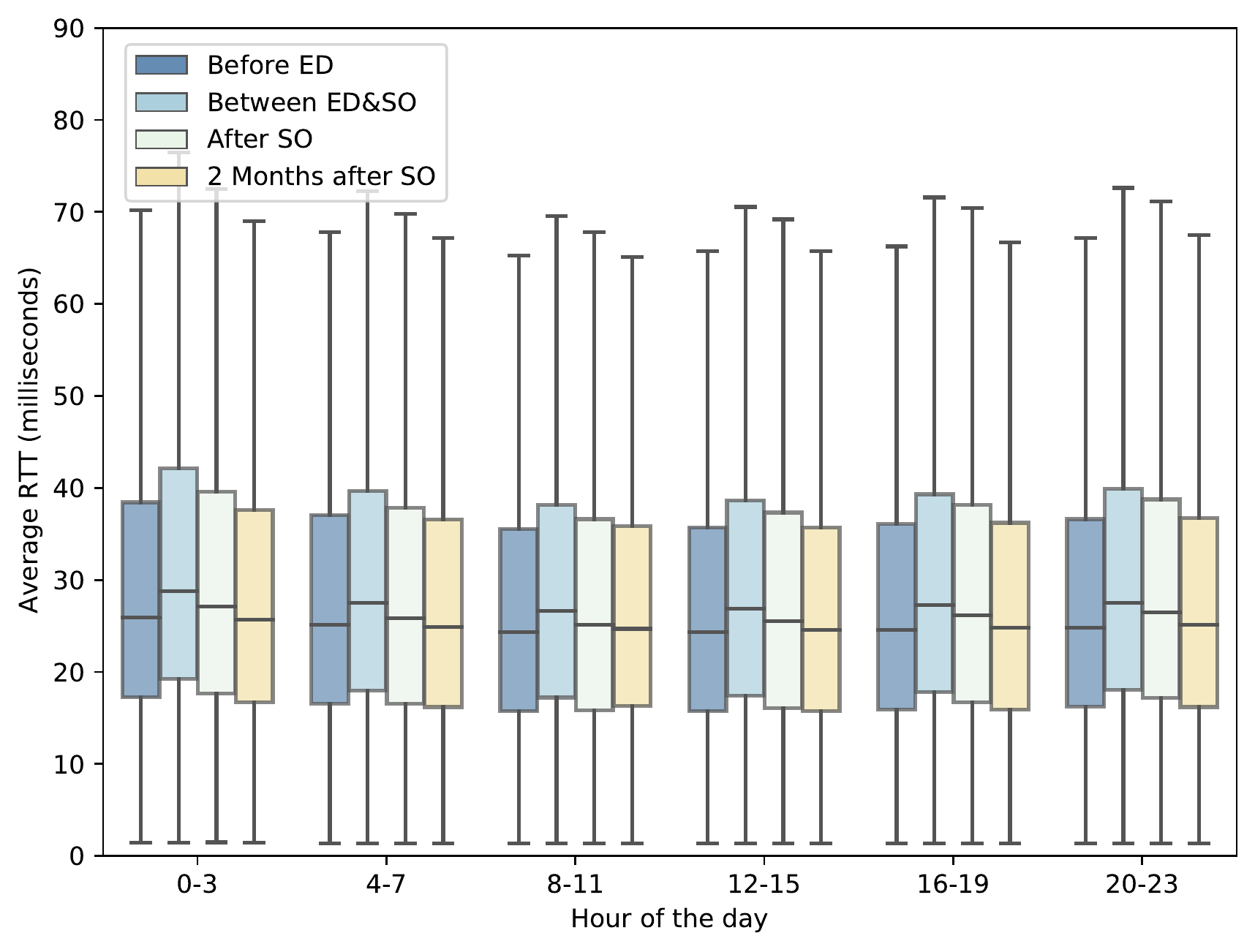}%
    }\hfil
    \subfloat[Throughput.\label{fig:hourly_T}]{%
      \includegraphics[width=0.5\textwidth]{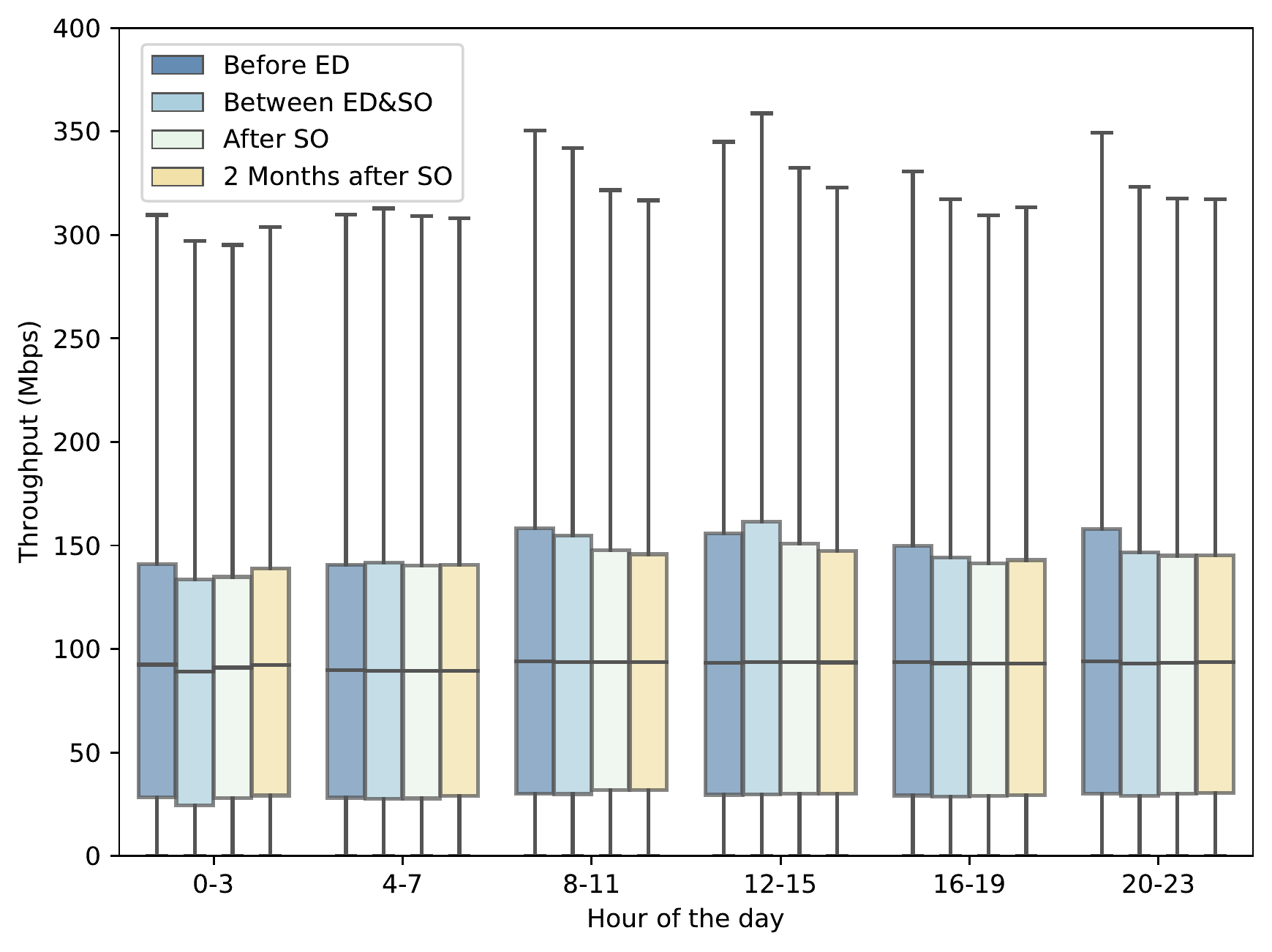}%
    }
    
    \caption{Changes in latency and throughput before and after the lockdown. ED
      means ``Emergency is declared'' SO means ``Stay-at-home Ordered''.}
    \label{fig:time}
     \end{figure*}

    \section{Longitudinal Latency Evolution for 2018--2019 (Previous Year)}
    This section provides a basis for performance comparison in Section ~\ref{sec:performance}. 
    Following the same analysis, we choose the exact same time period in the previous
    year (i.e., late 2018 to mid-2019) in the United States. We compute the
    average latency per-Whitebox per-day, and subsequently explore distributions
    across Whiteboxes for each ISP.
    
     \paragraph{Longitudinal evolution of aggregate, average round-trip latency.}
     Figure~\ref{fig:daily2} shows the aggregate average latency per-Whitebox per-day.
     The previous year has an overall latency of about 6ms lower than 2020. We observe 
     that the latency keeps stable until the end of April, where a deviation of 
     about 2ms is shown. The rate of increase is of about 10\%, echoing similar effects
     around lockdown. 
      
\begin{figure*}[h!]
    \centering
    \subfloat[95th percentile of ISP latency (Group 1)\label{fig:ISP_19_G1_95}]{%
        \includegraphics[width=0.7\textwidth]{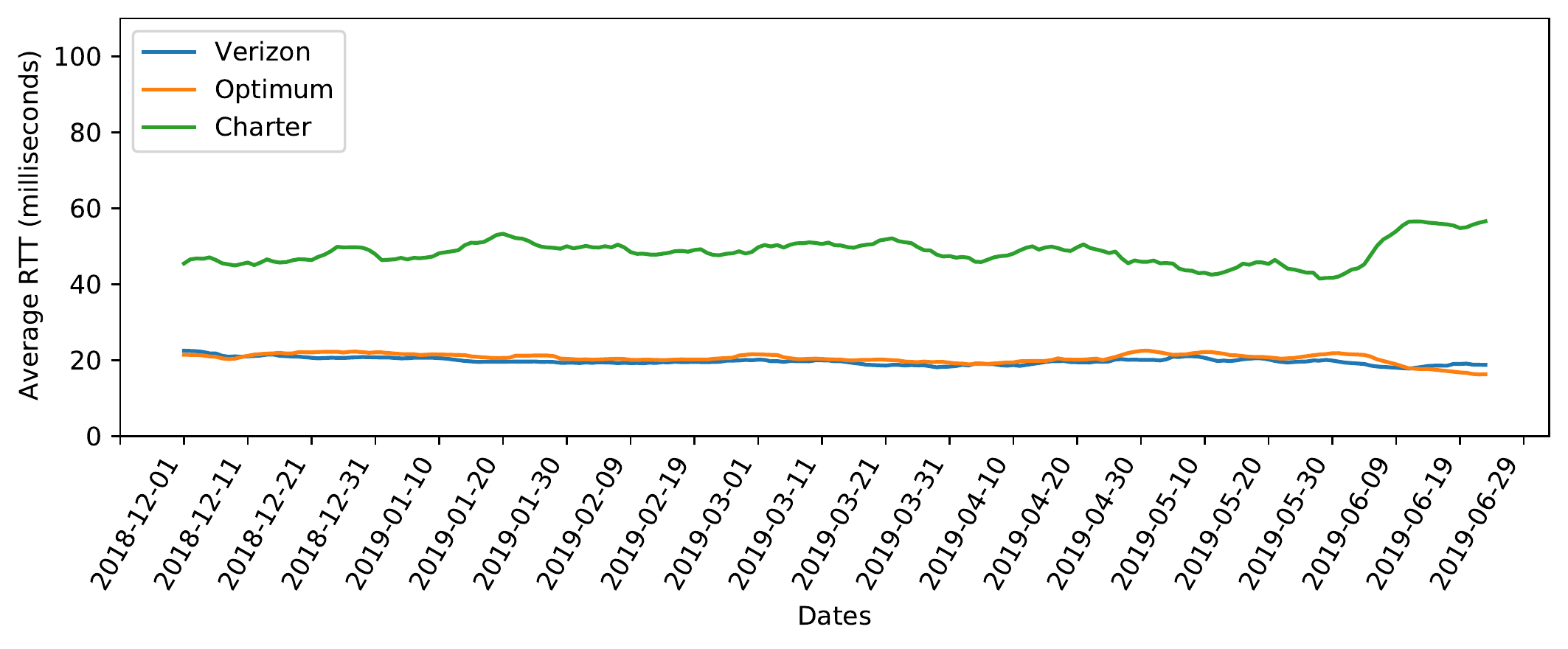}%
    }\hfil

    \subfloat[95th percentile of ISP latency (Group 2)\label{fig:ISP_19_G2_95}]{%
        \includegraphics[width=0.7\textwidth]{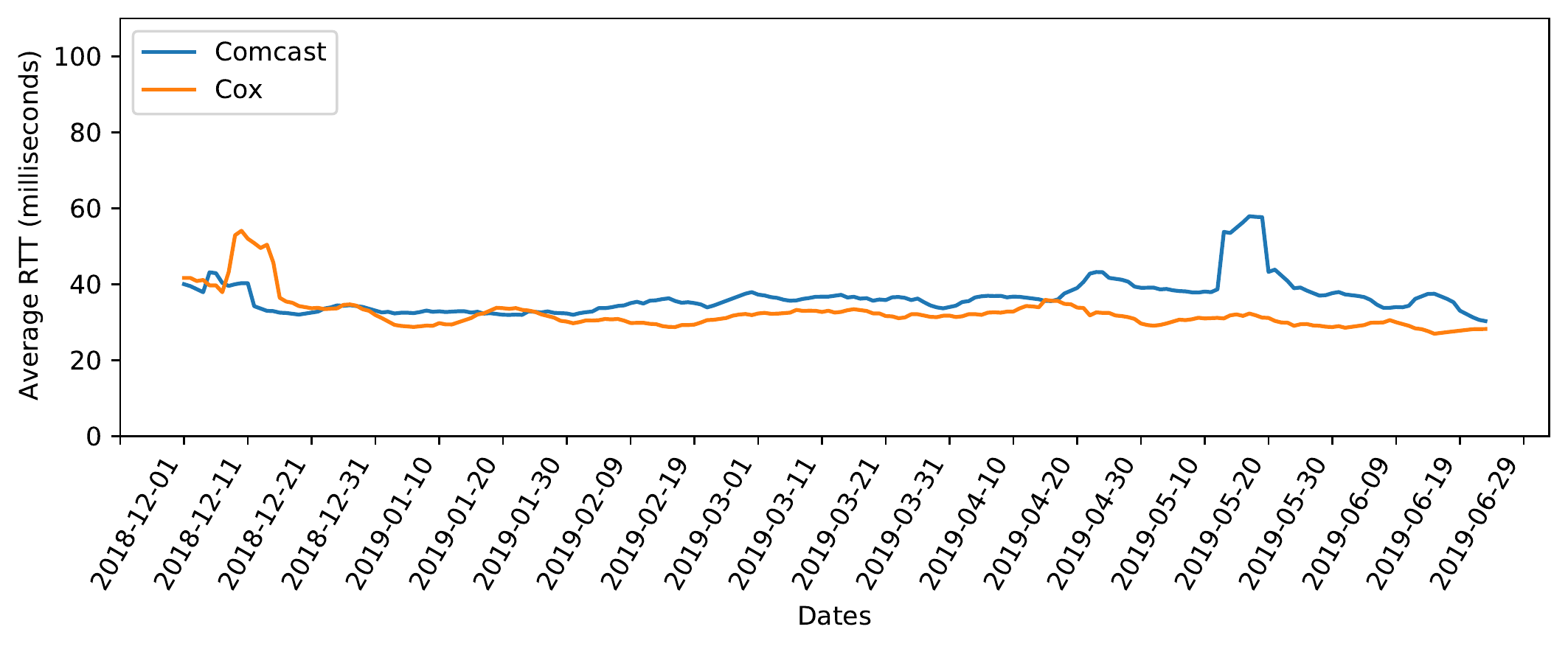}%
    }\hfil

    \subfloat[95th percentile of ISP latency with unstable changes (Group 3)\label{fig:ISP_19_G3_95}]{%
        \includegraphics[width=0.7\textwidth]{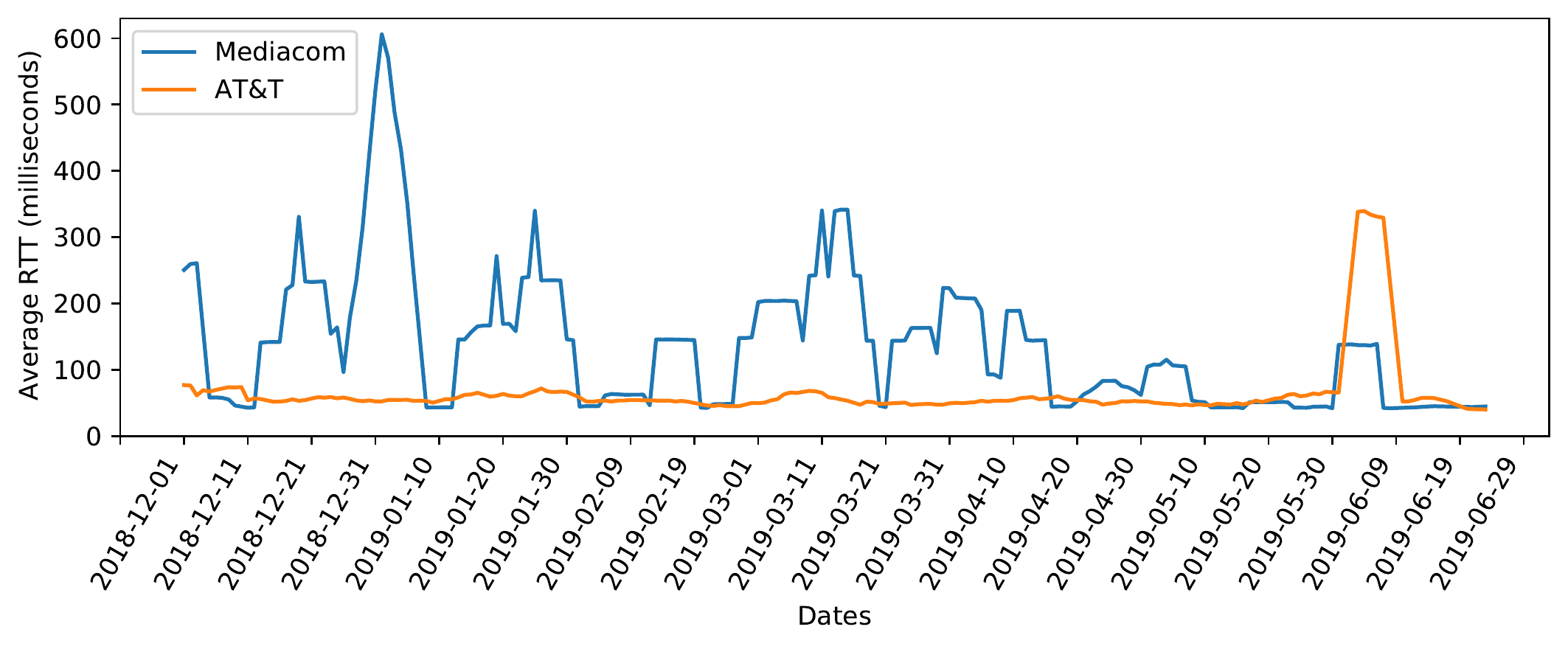}%
    }\hfil

    \caption{Latency (95th percentile) for different ISPs.}
    \label{fig:ISP_19_latency_95}

\end{figure*}

\begin{figure*}[t!]
    \centering
    \subfloat[99th percentile of ISP latency (Group 1)\label{fig:ISP_19_G1_99}]{%
        \includegraphics[width=0.7\textwidth]{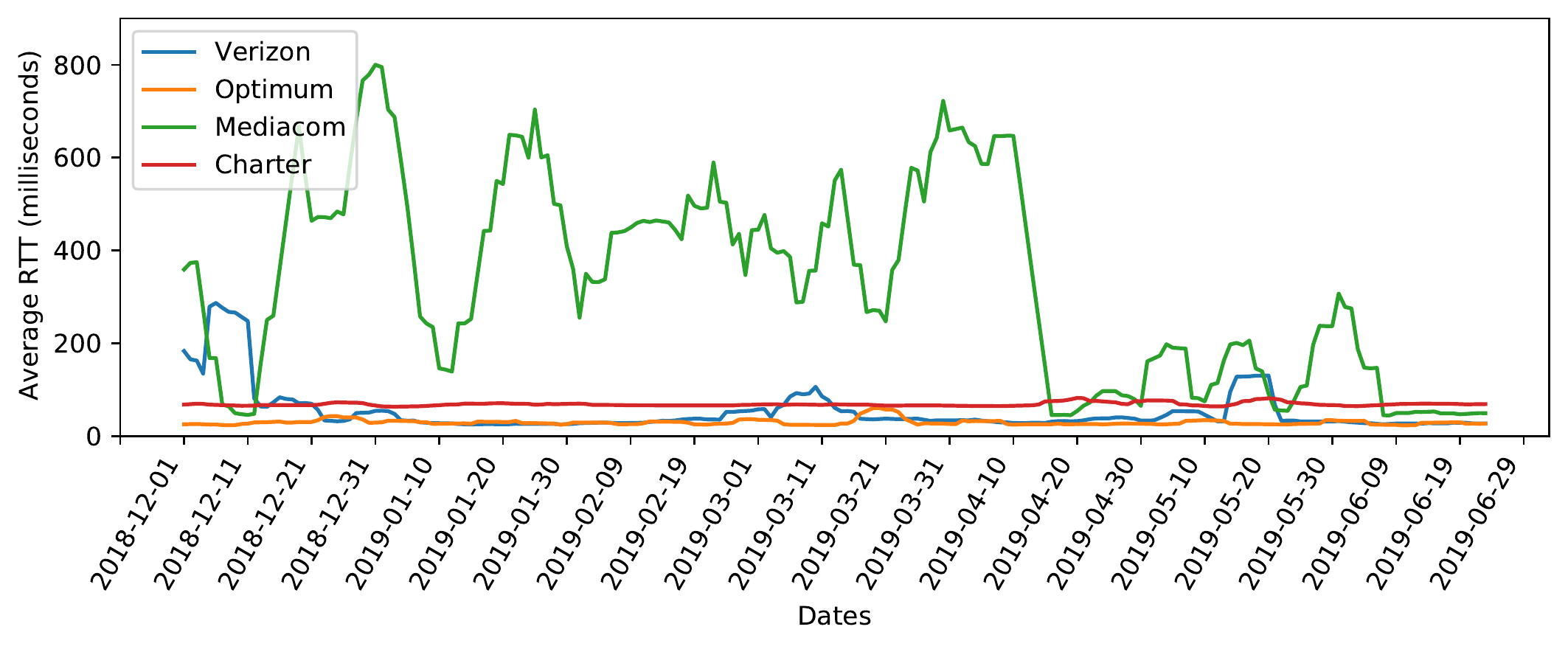}%
    }\hfil

    \subfloat[99th percentile of ISP latency (Group 2)\label{fig:ISP_19_G2_99}]{%
        \includegraphics[width=0.7\textwidth]{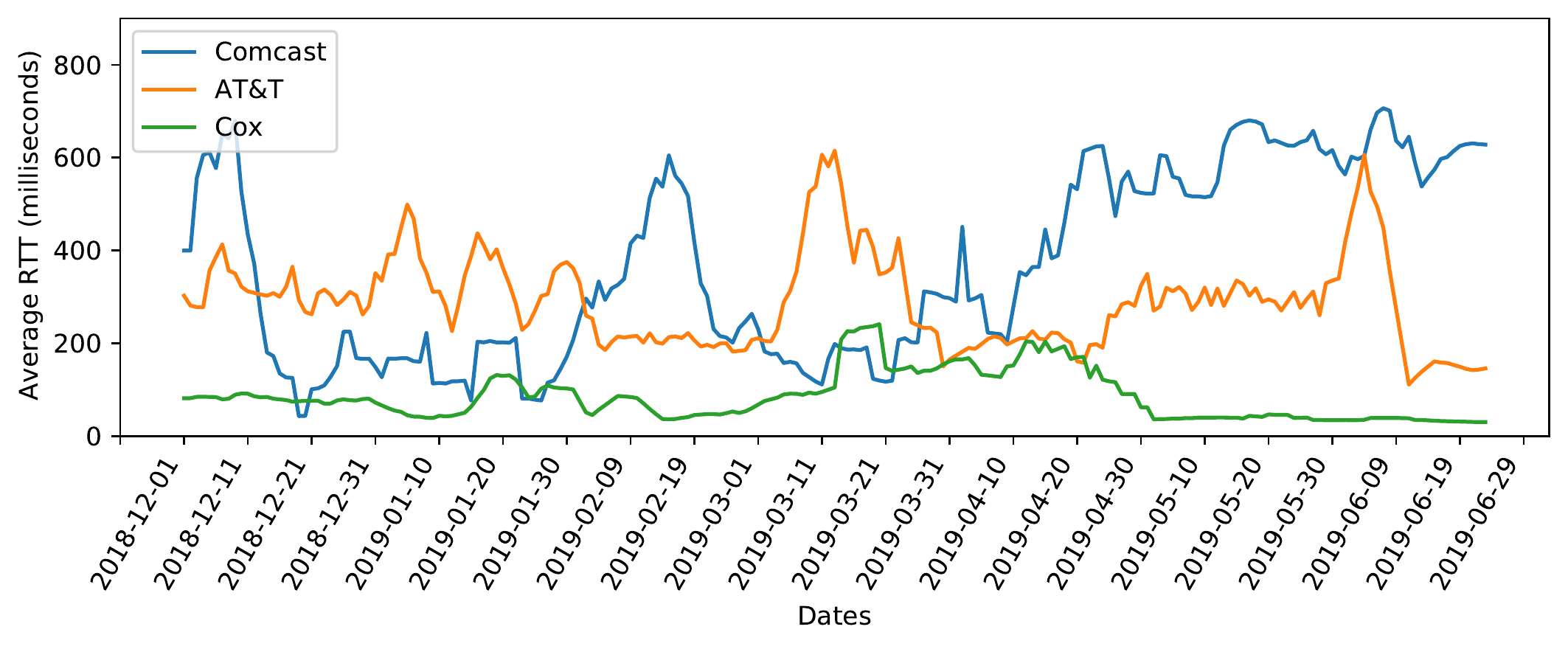}%
    }\hfil

    \caption{Latency (99th percentile) for different ISPs.}
    \label{fig:ISP_19_latency_99}

\end{figure*}

    \paragraph{Longitudinal evolution of per-ISP latencies.}
    We further break the aggregate results into the granularity of ISPs. We report both
    95th and 99th percentile latencies here. Note that in the 95th percentile plot, we 
    show the groups differently, mainly because of major differences of latency for 
    Mediacom and AT\&T. From Figure~\ref{fig:ISP_19_latency_95},
    we find that the majority of ISPs performed stably, while Mediacom has a large variance
    in the average RTT. They both have a tail that contributes to what we observed 
    in~\ref{fig:daily2}. Figure~\ref{fig:ISP_19_latency_99} is grouped the same as 
    Figure~\ref{fig:ISP_latency_99}, which shows that for certain ISPs, they experience
    similar deviations during similar periods of different years.

\section{Throughput-latency relationship}
    We put a supplementary figure referred to in Section~\ref{sec:performance} in this appendix.
    It shows the distributional changes in latency and throughput on a 4-hour basis. Detailed
    explanations are in the main text. 


    
 
